\documentclass[11pt]{article}
\usepackage{cospar}
\usepackage{graphicx}
\newcommand{\hb}{H{\sc $\beta$}}

\newcommand{\mgiil}{Mg\,{\sc ii}\,$\lambda$2798}

\newcommand{\feii}{Fe\,{\sc ii}}

\newcommand{\mnras}{\it MNRAS}
\newcommand{\apj}{{\it Astrophys. J.}}
\newcommand{\aj}{{\it Astro. J.}}
\newcommand{\aap}{{\it Astro. Astrophys.}}
\newcommand{\apjl}{{\it Astrophys. J. Lett.}}
\newcommand{\apjs}{{\it Astrophys. J. Supp.}}
\def\la{\mathrel{\hbox{\rlap{\hbox{\lower4pt\hbox{$\sim$}}}\hbox{$<$}}}}
\def\ga{\mathrel{\hbox{\rlap{\hbox{\lower4pt\hbox{$\sim$}}}\hbox{$>$}}}}

\begin{document}

% declarations for front matter
\title{INDICATORS OF BLACK HOLE MASS AND EDDINGTON ACCRETION RATIO FROM QSO X-RAY AND UV SPECTRA}

\author{Beverley J. Wills, and Zhaohui Shang\\
\vspace*{3mm}
{\it The University of Texas at Austin, 1 University Station C1400, Austin, 
TX 78712-0259, USA}
}

% typeset front matter
\maketitle

\begin{abstract}
The evolution of luminous QSOs is linked to the evolution of
massive galaxies.  We know this because the relic black-holes found
locally have masses dependent on the properties of the host galaxy's bulge.
An important way to explore this evolution would be to measure
dependences of black hole masses and Eddington accretion ratios over a range of
redshifts, i.e., with cosmological age.
For low redshift QSOs (and their lower luminosity Seyfert galaxy
counterparts) it has been possible to infer black hole masses from
the luminosities and velocity dispersions of their host-galaxy bulges.
These masses agree with those virial black hole masses calculated from
the Doppler widths of the broad H$\beta$ emission lines.  The
latter method can then be extended to more distant and luminous
QSOs, up to redshifts of 0.6 with ground-based optical observations.
We discuss ways to extend these explorations to higher redshifts -- up
to $\sim$3 using the widths of QSOs' broad UV emission lines,
and in principle, and to redshifts near 4 from ground-based
infrared observations of rest-frame H$\beta$ at 2.5$\mu$m.
We discuss the possibility of investigating the accretion history
of the higher redshift QSOs using measures of Eddington accretion
ratio -- the soft X-ray spectral index and the eigenvectors of
Principal Components Analyses of QSOs' UV emission-line spectra.
\end{abstract}

\section*{INTRODUCTION}
Black holes appear to be ubiquitous in the nuclei of nearby galaxies
(Magorrian et al. 1998)
with the black hole mass directly proportional to the bulge luminosity or to
$\sigma^4$, where $\sigma$ is the bulge velocity dispersion 
of the host galaxy (Kormendy and Gebhardt 2001, Tremaine et al. 2002).
The space density of these local supermassive
black holes is similar to that of QSOs at the heyday of their evolution
at redshifts of 2 -- 3.  By the present epochs, these QSOs have faded,
leaving behind the relic black holes found in local galaxies (Soltan 1982,
Fabian and Iwasawa 1999).  This inference,
together with the black-hole mass -- $\sigma$ relationship, shows that
the evolution of QSOs and galaxies is intimately related (e.g., Fabian 1999).
Measurement of black hole mass and Eddington accretion ratio of QSOs over
a wide redshift range, i.e., with cosmological epoch, would allow us to 
explore the accretion history of
QSOs, hence the evolution of galaxies.

In nearby, less-luminous active galaxies and QSOs it is difficult, but
possible, to measure the bulge mass via bulge luminosities and stellar
velocity dispersions.  This has allowed the calibration of a virial method
of measuring black hole masses using the broad H$\beta$ emission line of
active galaxies and nearby QSOs (Laor 1998, Gebhardt et al. 2000).  
In this method the H$\beta$ emission-line
gas within $\sim$1 parsec of the nucleus is assumed to move under the
gravitational influence of the central black hole.  Thus the H$\beta$
full-width at half maximum intensity (FWHM), together with the distance
of this gas inferred from reverberation mapping ($r \propto$ L$^{0.5}$,
where L represents the continuum luminosity -- e.g., Kaspi et al. 2000,
Vestergaard 2002, Maoz 2002), leads to a black hole mass:
$$ {\rm M}_{\rm BH} \sim {\rm FWHM}^2 {\rm L}^{0.5}$$
From this can be inferred an Eddington accretion ratio
$\propto {\rm L}/{\rm M}_{\rm BH}$ $\sim {\rm FWHM}^{-2} {\rm L}^{0.5}$.
Thus this H$\beta$ FWHM method can be used for higher luminosity and more
distant QSOs for which, in the
presence of the QSO nucleus, the host galaxy appears too small and dim
to measure bulge luminosity or velocity dispersion
(see Figure 4 in the paper by McLure and Dunlop 2002).

How can we extend these measurements to higher redshifts?

\section*{NEW WAYS TO MEASURE ${\rm M}_{\rm BH}$ AND ${\rm L}/{\rm M}_{\rm BH}$.}
There are several possibilities that we will discuss below.
\begin{enumerate}
\item{}{Near infrared observations of the rest-frame H$\beta$ emission
line can extend M$_{\rm BH}$ measurements to redshifts of 1.5 -- 2.5
(Yuan and Wills 2002a, 2002b).}
\item{}{It's been suggested that widths of emission lines of
Mg\,II$\lambda$2798 (McLure and Jarvis 2002) and C\,IV$\lambda$1549
(Vestergaard 2002)
could be used instead of H$\beta$, in principle extending the method to
redshifts of 1.9 and 4.2.   These authors show that the widths of these UV
lines can be used to estimate black hole masses, with a calibration based
on the \hb\ method (McLure and Jarvis' 2002 Figure 4; Vestergaard's 2002 Figure 4b).
However, the scatter is large.  Both UV lines, but especially
\mgiil, are seriously affected by \feii\ blends.  These \feii\ blends both
contaminate the broad emission lines and make continuum determination in the
region of the lines extremely unreliable (e.g., Vestergaard and Wilkes 2000).}
\item{}{At low redshifts, we have shown that X-ray spectra and optical-UV emission
line relationships are closely related to Eddington accretion ratio (Yuan and Wills
2002a, 2002b, Boroson 2002, Laor et al. 1997).  Emission-line relationships are probably
determined by the available fuel supply.  Thus Eddington accretion ratio and 
M$_{\rm BH}$ can be determined.}
\end{enumerate}

We illustrate how the relationships in 3. above can be employed to measure
Eddington accretion ratios, using results from the 18--22 QSO sample discussed
by Shang et al. (2002a, 2002b).

\vspace*{5mm}
\begin{minipage}{180mm}
\includegraphics[width=75mm]{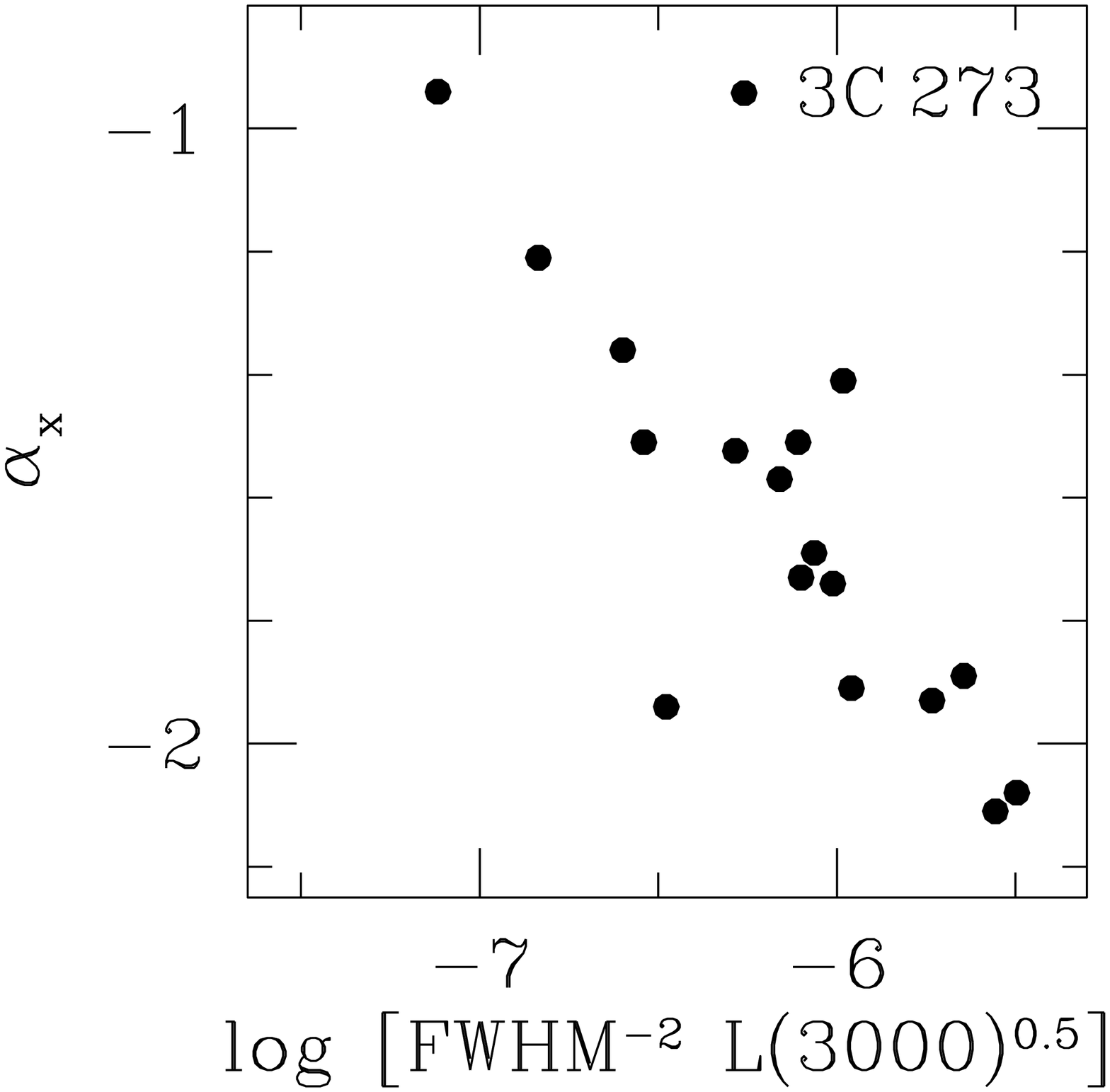}
\end{minipage}
\vspace*{10mm}

\begin{minipage}{180mm}
{\sf Fig.\ 1\ \ The dependence of soft X-ray spectral index $\alpha_{\rm x}$ on
the logarithm of FWHM$^{-2}$ L$^{0.5}$ (proportional to Eddington ratio).
}
\end{minipage}
%18 144 592 718

\subsection{X-ray Spectral Index and Eddington Ratio}

Figure 1 shows soft X-ray  spectral index $\alpha_{\rm x}$ (from Laor et al. 1997)
vs Eddington ratio, calculated from \hb\ widths as described above.
Measurement uncertainties on $\alpha_{\rm x}$ are typically $\pm 0.05$, and
the observational uncertainties on log [FWHM$^{-2}$ L(3000)$^{0.5}$] are
approximately $\pm 0.15$.  3C\,273 is a radio-loud core-dominant
QSO, with a strong jet contribution that flattens the X-ray spectrum.
Therefore it should not be included.  Without 3C\,273, the two-tailed 
probability P$_{\rm 2t}$ of a chance correlation is $<2$\,10$^{-5}$.  In principle
then, L/L$_{\rm Edd}$
can be determined from $\alpha_{\rm x}$.   At higher redshifts the rest frame
soft X-ray spectrum is unobservable, redshifted to photon energies where it is
absorbed by the interstellar medium in our Galaxy.  However there is a
correlation between hard and soft X-ray spectral indices (compare Fig.\,8 of
Boller, Brandt, and Fink 1996, with Fig.\,1 of Brandt, Mathur, and Elvis 1997).
The use of this relationship needs further understanding of
the complex X-ray spectra being revealed by Chandra and XMM.

\subsection{UV Emission Line Spectra and Eddington Ratio}

At reshifts $z \sim$1.9 -- 3.3, \hb\ is redshifted out of the optical atmospheric 
window, but UV rest wavelengths are observable.  So another approach is to use the
UV emission-line spectrum.

%Another approach is to use the UV emission line spectra to determine L/L$_{\rm Edd}$,
%extending the methods to $z \sim$1.9 -- 3.3.

One way is via a Principal Components Analysis of directly measured parameters.  
The method finds independent principal components, which are linear combinations 
of these measured input parameters.  The first principal component is the linear 
combination that accounts for most of the spectrum-to-spectrum variance, the second
principal component, the one that accounts for the next most variance, etc.
The method is explained and described by Francis and Wills (1999), and Wills et al.
(1999),
who apply the technique to the low redshift sample discussed here. 
Here we demonstrate, using our low redshift QSO sample, that the UV spectra
(observed with the Faint Object Spectrograph on the Hubble Space Telescope) can
predict the Eddington ratio determined from \hb\ FWHM.

% Only the UV spectrum can
%be observed at optical observed wavelengths, so here we are interested in how well
%the UV spectrum alone can predict the Eddington accretion ratio.

With just the UV emission-line strengths, ratios
and widths as input (the data are tabulated by Francis and Wills 1999), Principal 
Components Analysis reveals a first principal component, UV\,PC1, accounting for
48\% of the spectrum-to-spectrum variance.  This linear
combination of the input variables depends on L/L$_{\rm Edd}$ (Figure 2).

A related Principal Components Analysis that uses the whole spectrum was just 
described by Shang et al. (2002b).   
Each QSO spectrum is represented by the flux in small wavelength bins along the
entire spectrum.  Instead of measured line and continuum parameters, this spectral
PCA uses these binned fluxes as input parameters.  In an analysis of just the
UV spectrum (rest wavelengths 1171\AA -- 2100\AA), about half of the 
spectrum-to-spectrum variance is described by the first principal component, and
this linear combination of variables is dependent on luminosity (the Baldwin
effect), but independent of Eddington ratio.  The second most important linear
combination (the Second Principal Component, UV\,SPC2), accounting for 20\% of the 
spectrum-to-spectrum variance, is dependent on Eddington ratio (Figure 3).

Thus these linear combinations of input variables -- using either PCA or spectral PCA --
can be used to predict the Eddington accretion ratio.  Figure 4 shows, as
expected, the good correlation between the two linear combinations from PCA (UV\,PC1)
and from spectral PCA (UV\,SPC2).
The UV emission line principal components also agree well with the X-ray
predictor of Eddington accretion ratio (Figure 5).

\section{CONCLUSIONS}

At low redshift ($\la 0.6$),  QSO X-ray spectral index and UV emission line spectra
can predict the Eddington ratio, with small scatter.  Given an appropriate luminosity
one can therefore also derive M$_{\rm BH}$.   Spectral indicators can be observed at
higher redshift, thus, in principle, enabling the investigation of the evolution of 
Eddington accretion ratio and M$_{\rm BH}$ at epochs important for the growth of
black holes and galaxy spheroids.  Further investigation is needed to apply these
methods at higher redshift.  These relationships are empirical, and may evolve with
cosmic epoch.  We need to understand more clearly the physical basis of these
relationships.  At higher redshifts, only harder X-rays are available.  Chandra and
XMM spectroscopy may help us understand why the broad band spectral index shows these
dependences.  The UV emission line indicators may be a function of how the gas is
illuminated -- the geometry, the ionizing continuum.  They may also be indicators
of the available fuel supply -- and therefore only indirectly related to Eddington ratio.
One promising indication that these emission line properties are meaningful at redshifts
up to 2.5, is the demonstration that emission line properties in the redshifted
\hb\ region depend on Eddington accretion ratio, as derived from \hb\ FWHM (Yuan and
Wills 2002a, 2002b).
The existence of clear correlations at low redshift between Eddington ratio, 
M$_{\rm BH}$, optical-UV emission line properties, and soft Xray spectrum is, in
any case, giving us important information about the powering of QSOs' central
engines.   Whether these relationships are the same at high redshift (or luminosity),
they still hold important information about the evolution of the accretion process.

\begin{minipage}{80mm}
\includegraphics[width=75mm]{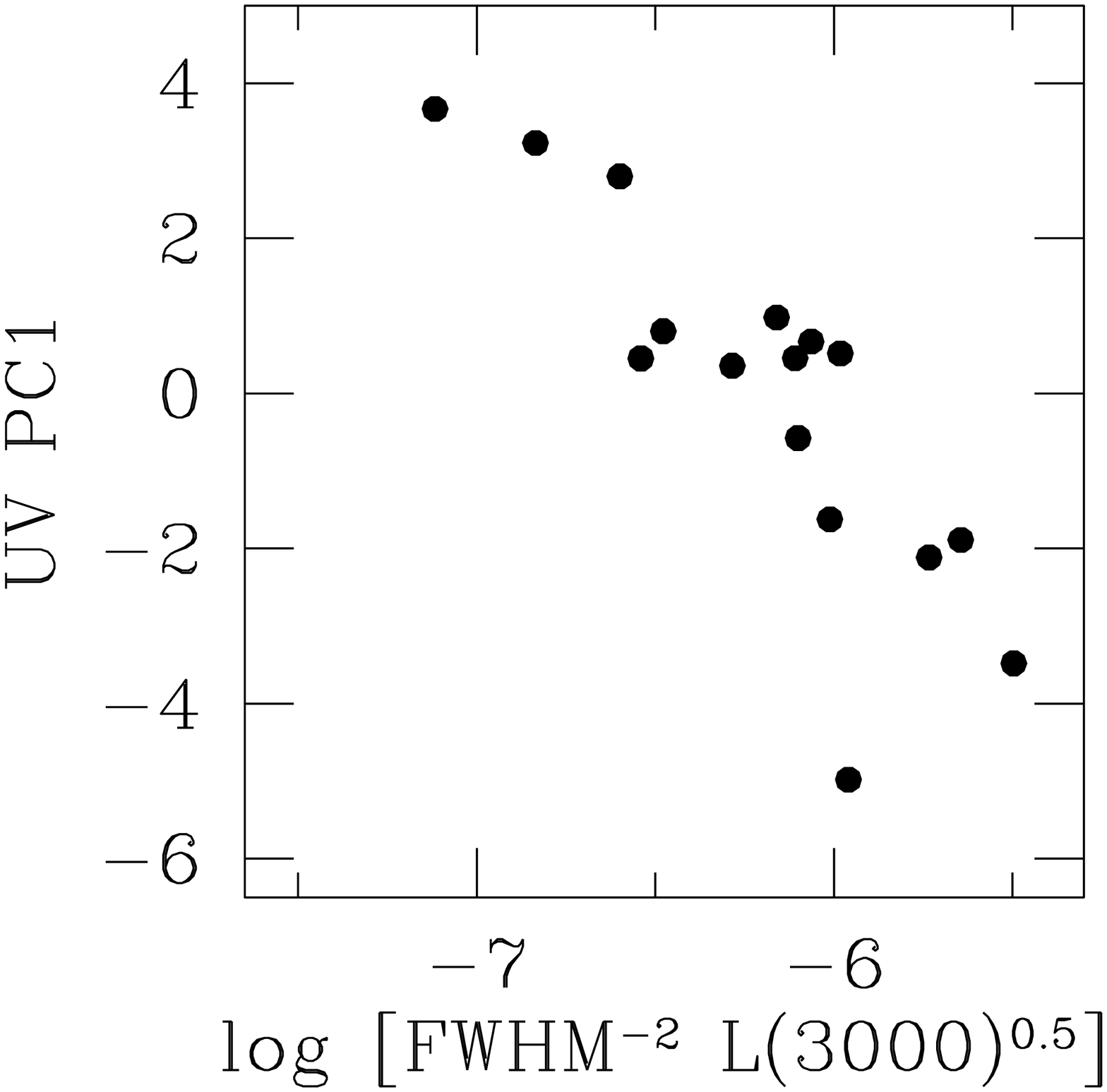}
\end{minipage}
\hfil\hspace{\fill}
\begin{minipage}{80mm}
\includegraphics[width=75mm]{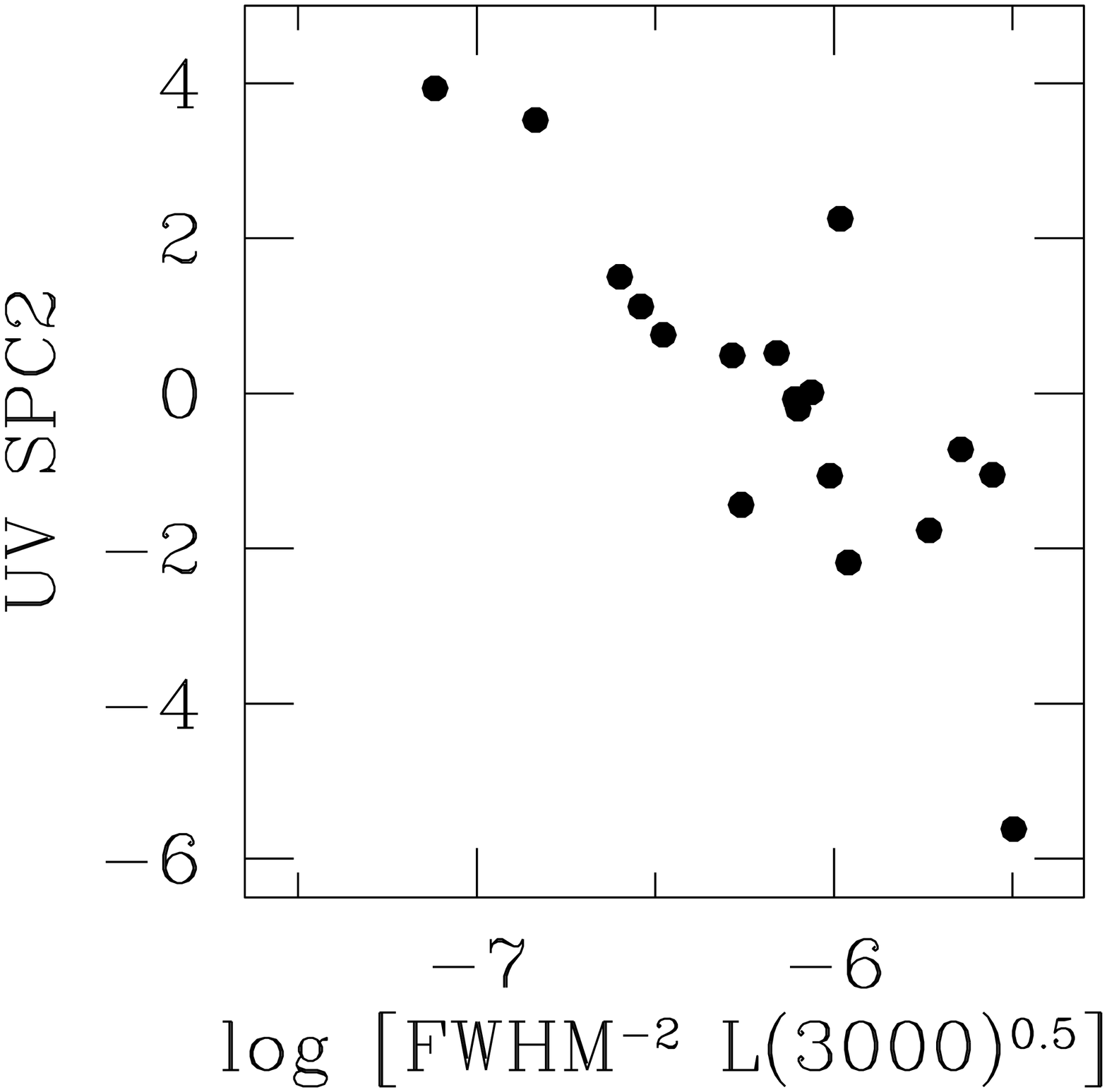}
\end{minipage}
\begin{minipage}{80mm}
{\sf Fig.\ 2.\ \ The dependence of Principal Component 1, derived from a PCA
of direct emission line measurements, on
the logarithm of FWHM$^{-2}$ L$^{0.5}$ (proportional to Eddington ratio).
P$_{\rm 2t} < 2 . 10^{-5}$.
}
\end{minipage}
\hfil\hspace{\fill}
\begin{minipage}{80mm}
{\sf Fig.\ 3.\ \ The dependence of Spectral Principal Component 2,
from an SPCA of the UV emission line spectra, on
the logarithm of FWHM$^{-2}$ L$^{0.5}$ (proportional to Eddington ratio).
P$_{\rm 2t} < 6 . 10^{-5}$.
}
\end{minipage}

\vspace*{2mm}
\begin{minipage}{80mm}
\includegraphics[width=75mm]{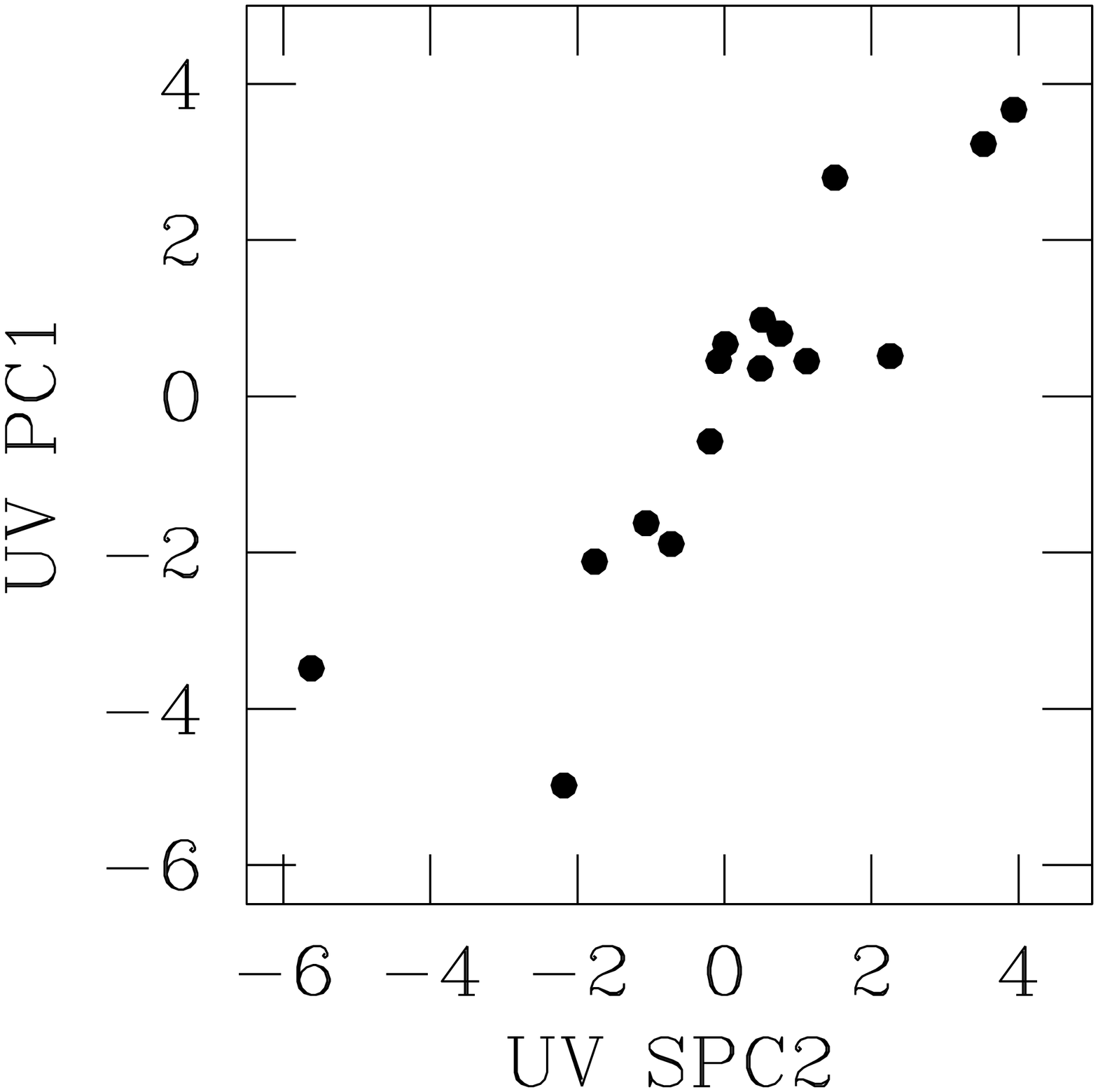}
\end{minipage}
\hfil\hspace{\fill}
\begin{minipage}{80mm}
\includegraphics[width=75mm]{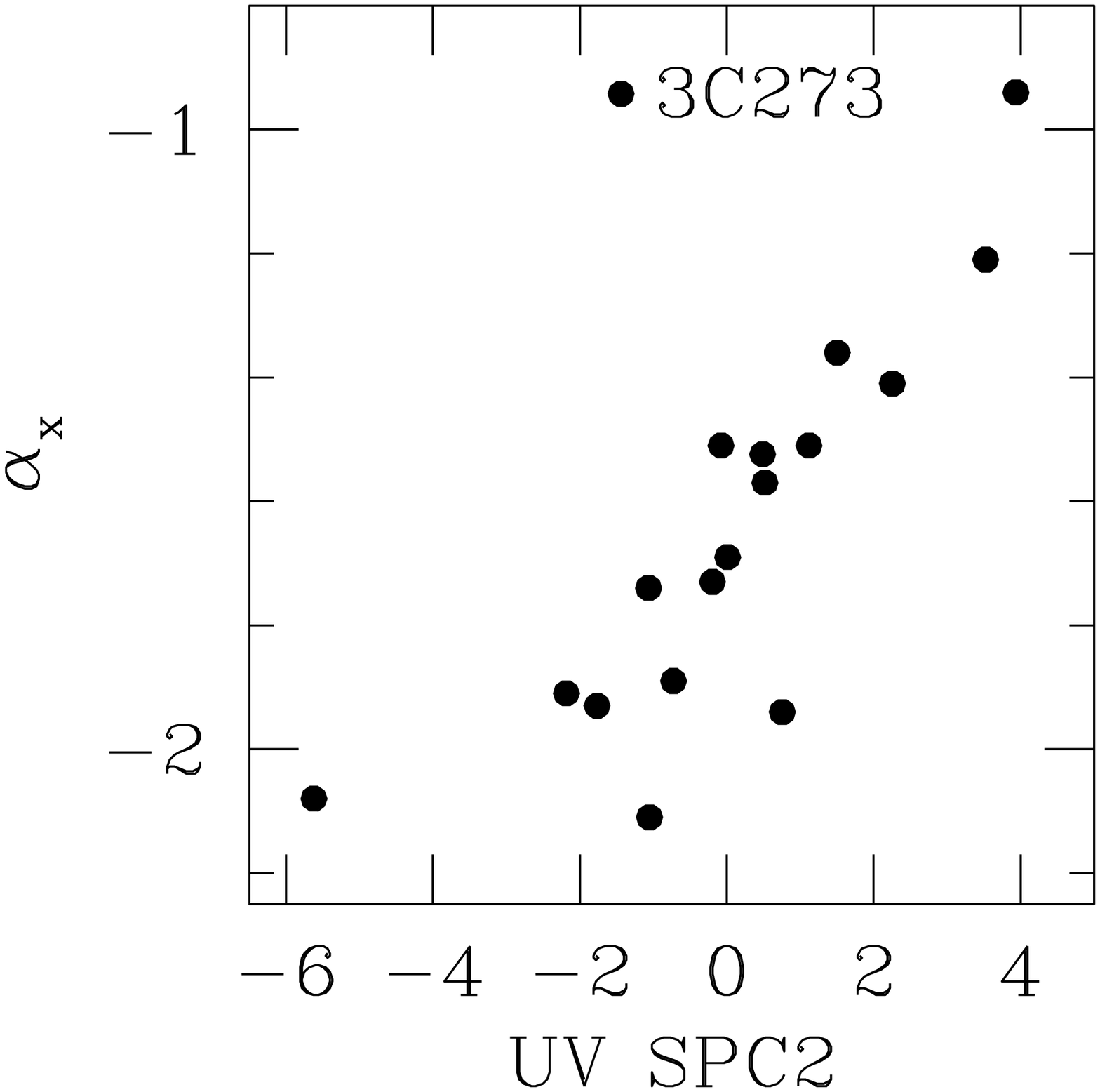}
\end{minipage}

\begin{minipage}{80mm}
{\sf Fig.\ 4.\ \ The dependence of principal component 1, derived from a
PCA  of direct measurements of UV emission lines, on
Spectral Principal Component 2, derived
from a spectral PCA of the UV emission line spectra.
P$_{\rm 2t} < 2 . 10^{-6}$.
}
\end{minipage}
\hfil\hspace{\fill}
\begin{minipage}{80mm}
{\sf Fig.\ 5.\ \ The dependence of soft X-ray spectral index $\alpha_{\rm x}$ on
Spectral Principal Component 2,
from a spectral PCA of the UV emission line spectra.
P$_{\rm 2t} < 2 . 10^{-5}$.
}
\end{minipage}

\vspace*{8mm}

\section*{ACKNOWLEDGEMENTS}

We thank Michael Yuan, Niel Brandt, and Sarah Gallagher for useful and interesting
discussions, and Derek Wills for a critical reading of the manuscript.
BJW is supported by the U.S. National Science Foundation, through grant
no. AST-0206261, and through NASA LTSA grant NAG5-3431.

\vspace*{10pt}

\noindent E-mail address of B. Wills: bev@astro.as.utexas.edu

\noindent Manuscript received 19 October 2002; revised \hspace{3cm} ;accepted

\end{document}